\documentstyle[preprint,seceq]{jpsj}%[preprint]

\newcommand{\vecvar}[1]{\mbox{\boldmath$#1$}}

\def\beq{\begin{equation}} \def\eeq{\end{equation}}
\def\bseq{\begin{subequations}} \def\eseq{\end{subequations}}
\def\bea{\begin{eqnarray}} \def\eea{\end{eqnarray}}
\let\nn=\nonumber
\def\beann{\begin{eqnarray*}} \def\eeann{\end{eqnarray*}}

\let\a=\alpha   \let\de=\delta
\let\ep=\varepsilon \let\z=\zeta

\newcommand{\poisson}[2]{\{#1\hspace{2.0pt}
 \mbox{\raisebox{2pt}{$\otimes$}}\hspace{-5.85pt}
 \mbox{\raisebox{-3pt}{,}}\hspace{5pt}#2\}}

 \newcommand{\eins}{\mbox{1 \hspace{-10.3pt} I}}

 \def\0{\over } \def\1{\vec }     \def\2{{1\over2}} \def\4{{1\over4}}
 \def\5{\bar }  \def\6{\partial } \def\7#1{{#1}\llap{/}}

 \def\<{\langle } \def\>{\rangle }

 \def\i{{\rm i}} \def\tr{\mbox{tr}} 
  \def\det{\mbox{det}} \def\sgn{\mbox{sgn}}

 \def\d{{\rm d}}

 \def\e{{\rm e}}

  \def\sech{\mbox{\,sech}}

\title{The Coupled Modified Korteweg-de Vries Equations}

\author{Takayuki {\sc Tsuchida}\footnote{E-mail:
 tsuchida@monet.phys.s.u-tokyo.ac.jp} and Miki {\sc Wadati}}

\inst{Department of Physics, Graduate School of Science, \\
 University of Tokyo,\\
 Hongo 7-3-1 Bunkyo-ku, Tokyo 113-0033}

\recdate{December 15 1997}

\abst{Generalization of the modified KdV
equation to a multi-component system, that is expressed by
$ \frac{\6 u_i}{\6 t} + 6 \bigl( \sum_{j,k=0}^{M-1} C_{jk} u_j u_k \bigr)
 \frac{\6 u_i}{\6 x} + \frac{\6^3 u_{i}}{\6 x^3} =0 ,
 \hspace{3mm} i=0, 1, \cdots, M-1 $, is studied.
 We apply a new extended version of the inverse scattering method to
  this system. It is shown that this system has an infinite number of 
 conservation laws and multi-soliton solutions. Further, the 
initial value problem of the model is solved.}

\kword{modified KdV equation, multi-component system, 
 inverse scattering method, conservation law, initial value problem, 
 soliton solution}

\begin{document}
\sloppy
\maketitle

\section{Introduction}
Since the discovery of the inverse scattering method (ISM), it has been
 shown that the ISM is applicable to many soliton
equations~\cite{Ablowitz1,Ablowitz2}. 
Among the soliton equations, the modified Korteweg-de Vries 
(mKdV) equation has been studied extensively because of its
simplicity and physical
significance.~\cite{Miura,Wadati1,Wadati2,Hirota}. 
Generalization of $2$~$\times$~$2$ AKNS formulation to
 $3$~$\times$~$3$ or, more generally, $L$~$\times$~$L$ formulation
 has already been
done by a number of authors.
~\cite{Ablowitz2,Manakov,Caudrey,Beals,ZM,Yajima,Sasa} 

Generalization of the mKdV equation
 to a multi-component system or a matrix equation 
has been studied by some authors.~\cite{Yajima,Sasa,Athorne,Fordy}
 One example is a vector
version of the mKdV equation proposed by Yajima and Oikawa~\cite{Yajima}.
 Sasa and Satsuma~\cite{Sasa}
 solved the initial value problem of the system, and constructed
 multi-soliton solution. Another example is a matrix version of the 
mKdV equation studied by Athorne and Fordy.~\cite{Athorne}

Recently, Iwao and Hirota~\cite{Iwao} discussed a simple coupled
version of the modified KdV equation,
 \beq
 \frac{\6 u_i}{\6 t} + 6 \biggl( \sum_{j,k=0}^{M-1} C_{jk} u_j u_k \biggr)
 \frac{\6 u_i}{\6 x} + \frac{\6^3 u_{i}}{\6 x^3} =0 ,
 \hspace{10mm} i=0, 1, \cdots, M-1,
 \label{CMKdV}
 \eeq
where the constants $C_{jk}$ are set to be symmetric with respect to 
the subscripts, $C_{jk}=C_{kj}$, without any loss of generality. 
They obtained multi-soliton solution of this system with the condition 
$C_{jj}=0$. Hirota~\cite{Hirota3} also studied solutions of
 semi-discrete version of the model. 
We call eq. (\ref{CMKdV}) the coupled modified KdV (cmKdV) equations. 
The cmKdV equations for $M=1$,~$2$ have been solved by the ISM. 
However, it has not been known whether the cmKdV equations 
for $M \ge 3$ and their hierarchy can be solved by the ISM or not.

In this paper, we propose a matrix generalization of the ISM that 
includes matrix mKdV equation, 
matrix nonlinear Schr\"{o}dinger equation (NLS) equation 
and other integrable equations. 
This formulation also includes the cmKdV equations as a 
reduction of the matrix mKdV equation. 
It provides us a method to solve the initial value problem and 
obtain soliton solutions.

The outline of the paper is as follows. In \S 2, we introduce a Lax
 representation for the matrix mKdV equation and the matrix 
NLS equation. Further the reduction of the 
matrix mKdV equation to the cmKdV equations is given. In \S 3, we perform
 the ISM for the matrix mKdV equation and the matrix NLS
 equation. In \S 4, we cast the results in \S 3 into 
those for the cmKdV equations. The last section, \S 5, is devoted to the
concluding remarks.

 \section{General Formulation}
 \label{formulation}

\subsection{Normalization}
\label{normalization}

We introduce an $M$-component vector field $\vecvar{u}$ and a constant 
$M \times M$ matrix $G$,
\beq
\vecvar{u} = (u_0, u_1, \cdots, u_{M-1})^{T}, \hspace{2mm} G=(-C_{ij}),
\label{}
\eeq
where the symbol $T$ means the transposition. 
Using this notation, a system of the cmKdV equations (\ref{CMKdV})
 is expressed as
\beq
\vecvar{u}_t - 6(\,\vecvar{u}\hspace{0.2mm}^{T}
 G \vecvar{u}\,)\, \vecvar{u}_x + \vecvar{u}_{xxx} = \vecvar{0}.
\label{u_eq}
\eeq
We assume that $G$ is a real symmetric and regular matrix
 in what follows. 
Because a real symmetric matrix is diagonalized by a real orthogonal
matrix, we can put
\bseq
\beq
G = P^{T} \Lambda P, \hspace{5mm} P^{T} P = P P^T = I, 
\eeq
\beq
\Lambda = 
{\rm diag} (\lambda_0, \cdots, \lambda_{M-1}), \hspace{2mm}
 \lambda_j \neq 0.
\label{}
\eeq
\eseq
Thus, defining  a new set of dependent variables 
$\vecvar{v} = (v_0, v_1, \cdots, v_{M-1})^{T}$ as
\beq
\vecvar{v} = P \vecvar{u},
\label{}
\eeq
eq.\ (\ref{u_eq}) is cast into
\beq
\vecvar{v}_t - 6(\,\vecvar{v}\hspace{0.2mm}^{T} \Lambda \vecvar{v}\,)\, \vecvar{v}_x + \vecvar{v}_{xxx} = \vecvar{0},
\label{}
\eeq
or more explicitly
\beq
 \frac{\6 v_i}{\6 t} - 6 \biggl( \sum_{j=0}^{M-1} \lambda_j v_j^2 \biggr)
 \frac{\6 v_i}{\6 x} + \frac{\6^3 v_{i}}{\6 x^3} =0 ,
 \hspace{2mm} \lambda_j \neq 0, \hspace{5mm} i=0, 1, \cdots, M-1.
 \label{CMKdV2}
\eeq
If we change a scale of $v_i$ by 
$\sqrt{|\lambda_i|} \cdot v_i$~$\to$~$v_i$,
 we finally obtain normalized cmKdV equations,
\beq
 \frac{\6 v_i}{\6 t} - 6 \biggl( \sum_{j=0}^{M-1} \ep_j v_j^2 \biggr)
 \frac{\6 v_i}{\6 x} + \frac{\6^3 v_{i}}{\6 x^3} =0 ,
 \hspace{3mm} \ep_j=\sgn(\lambda_j)=\pm 1,
 \hspace{4.5mm} i=0, 1, \cdots, M-1.
 \label{CMKdV3}
\eeq
The above equation (\ref{CMKdV3}) is more convenient than
eq.\ (\ref{CMKdV}) to perform the ISM. Thus, we mainly deal with
eq.\ (\ref{CMKdV3}) as the cmKdV equations in the following. In
addition, the dependent variables $\{ v_i \}$ are assumed to be real.

 \subsection{Lax pair}
 \label{Lax pair}
 We consider a set of auxiliary linear equations
 \beq
 \Psi_x = U \Psi, ~~\Psi_t = V \Psi.
 \label{scattering_problem1}
 \eeq
Here $\Psi$ is a $(p+q)$-component vector and $U$, $V$ are
 $(p+q) \times (p+q)$ matrices.
 The compatibility condition of eq.\ (\ref{scattering_problem1}) is 
 given by 
 \beq
 U_t -V_x + UV-VU = O.
 \label{Lax equation}
 \eeq
 We call $U$, $V$ Lax pair and eq.\ (\ref{Lax equation}) a
 zero-curvature condition, or Lax equation. 
 We introduce the following form of the Lax pair,
 \beq
 U
 =
 \i \z \left[
 \begin{array}{cc}
  -I_1  &  O \\
   O  &  I_2 \\
 \end{array}
 \right]
 +
 \left[
 \begin{array}{cc}
  O  &  Q \\
  R  &  O \\
 \end{array}
 \right],
 \label{U_form}
 \eeq
 \bea
 V
 &=&
 \i \z^3
 \left[
 \begin{array}{cc}
 -4I_1  &  O \\
  O  &  4I_2 \\
 \end{array}
 \right]
 + \z^2
 \left[
 \begin{array}{cc}
  O  &  4Q \\
  4R &  O  \\
 \end{array}
 \right]
 +\i \z
 \left[
 \begin{array}{cc}
  -2QR  & 2Q_x \\
  -2R_x & 2RQ  \\
 \end{array}
 \right]
 \nn \\
 && +
 \left[
 \begin{array}{cc}
  Q_x R-QR_x   &  -Q_{xx}+2QRQ \\
  -R_{xx}+2RQR &  R_x Q-RQ_x  \\
 \end{array}
 \right],
 \label{V_form}
 \eea
where $\z$ is the spectral parameter which does not depend on time, $\z_t=0$. 
$ I_1 $ and $ I_2 $ are respectively the $ p \times p $ and
$ q \times q $ unit matrices ; $Q$ is a $p \times q$ matrix 
(made up of $p$ rows and $q$ columns);
 $R$ is a $q \times p$ matrix. 

Substituting eqs.\ (\ref{U_form}) and (\ref{V_form}) into eq.\
 (\ref{Lax equation}), we get a set of matrix equations
\bseq
\beq
 Q_t + Q_{xxx} - 3Q_x RQ - 3QRQ_x = O,
 \label{}
 \eeq
\beq
 R_t + R_{xxx} - 3R_x QR - 3RQR_x = O.
 \label{}
\eeq
\label{mmKdV1}
\eseq
Suppose that 
$R$ is connected with the Hermitian conjugate of $Q$ by
\beq
R = \ep Q^{\dagger}, \hspace{3.5mm}\ep = \pm 1.
\label{R_eq_Q}
\eeq
Then, eq.\ (\ref{mmKdV1}) is reduced to
\beq
Q_t + Q_{xxx} -3\ep (Q_x Q^{\dagger} Q + Q Q^{\dagger} Q_x) = O,
\hspace{3.5mm}\ep = \pm 1.
\label{mmKdV2}
\eeq
If we restrict $Q$ to be a real matrix, eq.\ (\ref{mmKdV2}) becomes
equivalent to what Athorne and Fordy studied.~\cite{Athorne} 
We call eq.\ (\ref{mmKdV1}) or eq.\ (\ref{mmKdV2}) 
the matrix mKdV equation. 
We consider the ISM for eq.\ (\ref{mmKdV2}) with $\ep=-1$ 
in \S \ref{ISM}.

\subsection{Lax pair for the matrix NLS equation}
We employ another form of $V$ as
 \bea
 V
 &=&
 \i \z^2
 \left[
 \begin{array}{cc}
 -2I_1  &  O \\
  O  &  2I_2 \\
 \end{array}
 \right]
 + \z
 \left[
 \begin{array}{cc}
  O  &  2Q \\
  2R &  O  \\
 \end{array}
 \right]
 +\i 
 \left[
 \begin{array}{cc}
  -QR  & Q_x \\
  -R_x & RQ  \\
 \end{array}
 \right].
 \label{V_form2}
 \eea
Substituting eqs.\ (\ref{U_form}) and (\ref{V_form2}) into eq.\
 (\ref{Lax equation}), we get a set of matrix equations
\bseq
\beq
 \i Q_t + Q_{xx} - 2QRQ = O,
 \label{mNLS_1}
 \eeq
\beq
 \i R_t - R_{xx} + 2RQR = O.
 \label{mNLS_2}
\eeq
\label{mNLS1}
\eseq
Under the reduction
\beq
R = \ep Q^{\dagger}, \hspace{3.5mm}\ep = \pm 1,
\label{}
\eeq
eq.\ (\ref{mNLS1}) is cast into
\beq
\i Q_t + Q_{xx} - 2\ep Q Q^{\dagger} Q = O,
\hspace{3.5mm}\ep = \pm 1.
\label{mNLS2}
\eeq
We call eq.\ (\ref{mNLS1}) or eq.\ (\ref{mNLS2}) 
the matrix NLS equation.~\cite{ZM} 
By changing the time dependence of the ISM in \S \ref{ISM}, we can 
solve the initial value problem for this system (\ref{mNLS2}) with $\ep=-1$.

\subsection{Conservation laws}
\label{Conservation laws}

In this subsection, we present a systematic method to construct 
local conservation laws for the matrix mKdV equation and the matrix 
NLS equation. 
We start from a special class, $p=q=n$, 
of eq.\ (\ref{scattering_problem1}),
\beq
\left[
\begin{array}{c}
 \Psi_1  \\
 \Psi_2  \\
\end{array}
\right]_x
=
\left[
\begin{array}{cc}
 U_{11} & U_{12} \\
 U_{21} & U_{22} \\
\end{array}
\right]
\left[
\begin{array}{c}
 \Psi_1  \\
 \Psi_2  \\
\end{array}
\right],
\label{cons1}
\eeq
\beq
\left[
\begin{array}{c}
 \Psi_1  \\
 \Psi_2  \\
\end{array}
\right]_t
=
\left[
\begin{array}{cc}
 V_{11} & V_{12} \\
 V_{21} & V_{22} \\
\end{array}
\right]
\left[
\begin{array}{c}
 \Psi_1  \\
 \Psi_2  \\
\end{array}
\right],
\label{cons2}
\eeq
where all the entries in eqs.\ (\ref{cons1}) and (\ref{cons2}) are 
assumed to be $n \times n$ square matrices. 
The following is an extension of the method for the $n=1$ case.~
\cite{WSK} If we define a square matrix $\Gamma$ by
\beq
\Gamma \equiv \Psi_2 \Psi_1^{-1},
\label{Gamma}
\eeq
we can prove the following relations from eqs.\ (\ref{Lax equation}), 
(\ref{cons1}) and (\ref{cons2}),
\beq
\{ \tr (U_{12} \Gamma + U_{11}) \}_t = \{ \tr (V_{12}\Gamma + V_{11})\}_x,
\label{cons3}
\eeq
\beq
\Gamma_x = U_{21} + U_{22}\Gamma - \Gamma U_{11} - \Gamma U_{12} \Gamma .
\label{cons4}
\eeq
Equation (\ref{cons4}) is interpreted as the matrix Riccati equation. 
Assuming that $U$ is expressed as eq.\ (\ref{U_form}), we have
\beq
U_{11} = -\i \z I, 
\hspace{3mm}U_{22} = \i \z I, 
\hspace{3mm}U_{12} = Q, 
\hspace{3mm}U_{21} = R, 
\label{cons5}
\eeq
where $Q$ and $R$ are square matrices in this case. Then eqs.\ 
(\ref{cons3}) and (\ref{cons4}) are cast into the following 
equations,
\beq
\{ \tr \, (Q\Gamma) \}_t = \{ {\rm some \; function \; of}\; \Gamma, 
\hspace{1mm} Q, \hspace{1mm}R
\;  {\rm and}\;  \z \}_x,
\label{cons6}
\eeq
\beq
2\i \z Q\Gamma = -QR + Q(Q^{-1} \cdot Q\Gamma)_x + (Q\Gamma)^2 .
\label{cons7}
\eeq
Equation (\ref{cons6}) is nothing but a local conservation law. 
It shows that $\tr \hspace{0.3mm}(Q\Gamma)$ is a generating function of the 
conserved densities. We expand
 $Q\Gamma$ with respect to the spectral parameter 
$\z$ as follows,
\beq
Q\Gamma = \sum_{l=1}^{\infty} \frac{1}{(2\i \z)^l} F_l.
\label{cons8}
\eeq
Substituting eq.\ (\ref{cons8}) for eq.\ (\ref{cons7}),
 we obtain a recursion 
formula,
\beq
F_{l+1} = -\de_{l,0}QR + Q(Q^{-1}F_l)_x + \sum_{k=1}^{l-1}
F_k F_{l-k},
\hspace{5mm} l=0, 1, \cdots.
\label{F_formula}
\eeq
Each $\tr \, F_l$ is a conserved density for
 all positive integer $l$. 
Using the above formula (\ref{F_formula}), 
first four conserved densities are given by
\beq
F_1 = -QR,
\label{}
\eeq
\beq
\tr F_2 = \tr \{-QR_x\},
\label{}
\eeq
\beq
\tr F_3 = \tr \{-QR_{xx} + QRQR\},
\label{}
\eeq
\beq
\tr F_4 = \tr \{-QR_{xxx} + 2QR_x QR + QRQ_x R + 2QRQR_x\}.
\label{}
\eeq
It should be noted that all elements of $F_1=-QR$ are conserved 
densities for the matrix mKdV equation (\ref{mmKdV1}) and 
the matrix NLS equation (\ref{mNLS1}). 
This fact can be proved simply by a direct calculation.

\subsection{Hamiltonian structure of the matrix mKdV equation and the 
matrix NLS equation}
\label{Hamiltonian structure}

In this subsection, we consider the matrix mKdV equation and the matrix 
NLS equation with the condition that $Q$ and $R$ are $n \times n$ 
square matrices. 
The matrix mKdV equation (\ref{mmKdV1}) has the following Hamiltonian
structure. A set of the Hamiltonian and the Poisson bracket is 
\beq
H = \tr \int \{\i F_4\} \d x
= \tr \int 
%\biggl\{ -\i QR_{xxx} + \i QRQ_x R + \i QR_x QR + 2\i QRQR_x \biggr\} \d x,
\biggl\{-\i QR_{xxx}+\i \frac{3}{2}QR (QR_x - Q_x R) \biggr\} \d x,
\label{Hamiltonian}
\eeq
and
\bseq
\beq
\poisson{Q(x)}{Q(y)}
 = \poisson{R(x)}{R(y)} = O,
\label{PB1}
\eeq
\beq
\poisson{Q(x)}{R(y)} = \i \de(x-y) \Pi,
\label{PB2}
\eeq
\label{PB_mKdV}
\eseq
where $\poisson{X}{Y}_{kl}^{ij} = \{X_{ij}, Y_{kl}\}$ for
 matrices $X$, $Y$ and $\Pi$ denotes an $n^2 \times n^2$ permutation matrix. 
For the matrix NLS equation, the Hamiltonian is
\beq
H' = \tr \int \{ -F_2 \} \d x
= \tr \int 
\{ QR_{xx} - QR QR \} \d x,
\label{Hamiltonian2}
\eeq
instead of eq.\ (\ref{Hamiltonian}), while the Poisson bracket is 
the same. 
We can rewrite Poisson bracket between each element of $Q$ and $R$ 
explicitly as follows,
\bseq
\beq
\{Q_{ij}(x), Q_{kl}(y)\}
 = \{R_{ij}(x), R_{kl}(y)\} = 0,
\label{PB3}
\eeq
\beq
\{Q_{ij}(x), R_{kl}(y)\} = \i \de_{il} \de_{jk} \de(x-y)
 = \i \Pi^{ij}_{kl} \de(x-y) .
\label{PB4}
\eeq
\eseq
Indeed, we can check the following relations,
\beq
Q_t = \{Q, H\} = -Q_{xxx}+3Q_x RQ + 3QRQ_x,
\label{}
\eeq
\beq
R_t = \{R, H\} = -R_{xxx}+3R_x QR + 3RQR_x,
\label{}
\eeq
which are nothing but eq.\ (\ref{mmKdV1}). 
It shows that
 eq.\ (\ref{Hamiltonian}) and eq.\ (\ref{PB_mKdV}) are respectively 
the Hamiltonian and the Poisson bracket for the matrix mKdV equation. 
The same thing is true for the matrix NLS equation.

\subsection{r-matrix representation of the matrix mKdV equation and 
the matrix NLS equation}
\label{}

In \S \ref{Conservation laws}, we have shown that the 
 matrix mKdV equation (\ref{mmKdV1}) and the matrix NLS equation 
(\ref{mNLS1}) have an infinite number of
  conservation laws. In this subsection, we show that all the integrals
 of motion are in involution. In the following, we consider the 
 systems with an infinite interval and assume 
 the rapidly decreasing boundary conditions,
 \beq
 Q(x,t), \; R(x,t) 
 \rightarrow O \hspace{4mm} {\rm as}  \hspace{2mm}
 x\rightarrow \pm \infty.
 \label{boundary}
 \eeq

 If we define a classical r-matrix by
\beq
 {\rm r}(\z_1, \z_2) \equiv \frac{1}{2(\z_1 - \z_2)}
 \left[
 \begin{array}{cccc}
  \Pi &  &  &    \\
   & O &  \Pi &    \\
   &  \Pi & O &    \\
   &  &  &  \Pi   \\
 \end{array}
 \right],
 %=
 %\frac{1}{2\i( \z_1 - \z_2)}
 %\left[
 %\begin{array}{cccc}
 %  1 &  &  &    \\
 %  & 0 & 1 &    \\
 %  & 1 & 0 &    \\
 %  &  &  & 1   \\
 %\end{array}
 %\right]
 %\otimes \Pi,
 \label{}
 \eeq
 the following relation
 \beq
 \poisson{U(x;\z_1)}{U(y;\z_2)}
 = \de(x-y)\biggl[
{\rm r}(\z_1, \z_2),~
 U(x;\z_1)\otimes 
 \left[
 \begin{array}{cc}
  I & O \\
  O & I \\
 \end{array}
 \right]
 +
 \left[
 \begin{array}{cc}
  I & O \\
  O & I \\
 \end{array}
 \right]
 \otimes U(x;\z_2)
 \biggl],
 \label{r-representation}
 \eeq
 is satisfied. Here $[A,~B]\equiv AB-BA$ is a commutator and $U$ is 
 given by eq.\ (\ref{U_form}). By means of eq.\ (\ref{r-representation}), 
 we obtain a relation for transition matrices,~\cite{Korepin}
 \beq
 \poisson{T(x,y;\z_1)}{T(x,y;\z_2)}
 = [{\rm r}(\z_1, \z_2), 
 ~T(x,y;\z_1)\otimes T(x,y;\z_2)]
 \label{com1}
 \eeq
 where $T(x,y;\z)$ is the transition matrix defined by
 \beq
 T(x,y;\z) = {\rm P} \exp \biggl\{ \int_y^x U(z,\z) \d z\biggr\},
 \label{com2}
 \eeq
 with ${\rm P}$ being the path ordering. 
% which satisfies 
% \beq
% {\rm P} U(z_1, \z)U(z_2, \z)\cdots U(z_l, \z)
% = U(z_{i_1}, \z)U(z_{i_2}, \z)\cdots U(z_{i_l}, \z),
% \hspace{2mm} z_{i_1}\ge z_{i_2}\ge \cdots \ge z_{i_l} .
% \label{}
% \eeq 

 Taking the traces of both sides of eq.\ (\ref{com1}) , we get 
 \beq
 \{ \log \tau(\z_1), \log \tau (\z_2)\} = 0,
 \label{com3}
 \eeq
 where $\tau(\z)$ is defined by
 %\beq
 %\tau(\z) = {\rm tr}~ T(L,0;\z),
 %\label{}
 %\eeq
 %for the periodic case with a period $L$ and
 \beq
 \tau(\z) = {\rm tr}~ T(\infty,-\infty;\z),
 \label{}
 \eeq
 for the systems with an infinite interval. 
 Expanding (\ref{com3}) with respect to the spectral parameters 
 $\z_1$ and $\z_2$, we have the involutiveness of conserved quantities 
 $\{ J_n \}$,
 \beq
 \{ J_n, J_m \} = 0.
 \label{}
 \eeq
 This fact indicates the complete integrability of the matrix mKdV
equation and the matrix NLS equation.

 \subsection{Reduction of the Lax pair for the coupled modified KdV equations}
 \label{}
 In this subsection, we show a method to reduce the matrix mKdV equation to
 the cmKdV equations. We recursively 
 define $2^{m-1} \times 2^{m-1}$ matrices $Q^{(m)}$ and $R^{(m)}$ by
 \beq
 Q^{(1)} = \mu_0 v_0 + \i v_1, \hspace{5mm}
 R^{(1)} = \ep_1 (\mu_0 v_0 - \i v_1) ,
 \label{QR_def1}
 \eeq
 \beq
 Q^{(m+1)} 
 = 
 \left[
 \begin{array}{cc}
  Q^{(m)} &  -\ep_{2m+1}(\mu_{2m} v_{2m}+\i v_{2m+1}) I_{2^{m-1}} \\
  -(\mu_{2m} v_{2m}-\i v_{2m+1})I_{2^{m-1}} &  -R^{(m)} \\
 \end{array}
 \right],
 \label{QR_def2}
 \eeq
 \beq
 R^{(m+1)} 
 = 
 \left[
 \begin{array}{cc}
  R^{(m)} &  -\ep_{2m+1}(\mu_{2m} v_{2m}+\i v_{2m+1}) I_{2^{m-1}} \\
  -(\mu_{2m} v_{2m}-\i v_{2m+1})I_{2^{m-1}} &  -Q^{(m)} \\
 \end{array}
 \right].
 \label{QR_def3}
 \eeq
 Here $I_{2^{m-1}}$ is the $2^{m-1} \times 2^{m-1}$
 unit matrix. Each $\ep_i$ is a constant 
which is either $1$ or $-1$. 
$\mu_{2m}$ is a constant that satisfies
 \beq
 \mu_{2m}^2 = \frac{\ep_{2m}}{\ep_{2m+1}}=\ep_{2m}\ep_{2m+1}.
 \label{}
 \eeq
 For $Q^{(m)}$ and $R^{(m)}$ defined by eqs.\ (\ref{QR_def1})--(\ref{QR_def3}), 
 we can prove a simple relation,
 \beq
 Q^{(m)} R^{(m)} =  R^{(m)} Q^{(m)} = 
\sum_{j=0}^{2m-1} \ep_j v_j^2 \cdot I_{2^{m-1}},
 \label{}
 \eeq
 by the induction method. Then substituting $Q^{(m)}$ and $R^{(m)}$ for $Q$ and $R$ 
 in the matrix mKdV equation (\ref{mmKdV1}), we obtain
 \beq
  \frac{\6 v_i}{\6 t} - 6 \biggl( \sum_{j=0}^{M-1} \ep_j v_j^2 \biggr)
  \frac{\6 v_i}{\6 x} + \frac{\6^3 v_{i}}{\6 x^3} =0 ,
  \hspace{3mm} \ep_j=\pm 1, \hspace{7mm} i=0, 1, \cdots, M-1,
  \label{CMKdV4}
 \eeq
where we set $M = 2m$. 

In the following, we choose
 \beq
 \ep_i = -1, \hspace{5mm} i = 0, 1, \cdots, 2m-1, 
 %\hspace{4mm}
 \label{}
 \eeq
 \beq
 \mu_{2j} = 1, \hspace{5mm} j=0, 1, \cdots, m-1,
 \label{}
 \eeq
 and consider a self-focusing type of the 
cmKdV equations (\ref{CMKdV3}), that is,
 \beq
 \frac{\6 v_i}{\6 t} + 6 \biggl( \sum_{j=0}^{M-1} v_j^2 \biggr)
 \frac{\6 v_i}{\6 x} + \frac{\6^3 v_{i}}{\6 x^3} =0 ,
 \hspace{5mm} i=0, 1, \cdots, M-1 .
 \label{CMKdV5}
 \eeq
In this case, the recursion relations of $Q^{(m)}$ and $R^{(m)}$
 are
 \beq
 Q^{(1)} = v_0+ \i v_1, \hspace{5mm} R^{(1)} = -v_0+\i v_1
 \label{QR_def4}
 \eeq
 \beq
 Q^{(m+1)} 
 = 
 \left[
 \begin{array}{cc}
  Q^{(m)} &  (v_{2m}+\i v_{2m+1}) I_{2^{m-1}} \\
  -(v_{2m}-\i v_{2m+1})I_{2^{m-1}} &  -R^{(m)} \\
 \end{array}
 \right],
 \label{QR_def5}
 \eeq
 \beq
 R^{(m+1)} 
 = 
 \left[
 \begin{array}{cc}
  R^{(m)} &  (v_{2m}+\i v_{2m+1}) I_{2^{m-1}} \\
  -(v_{2m}-\i v_{2m+1})I_{2^{m-1}} &  -Q^{(m)} \\
 \end{array}
 \right].
 \label{QR_def6}
 \eeq
 We can show that a simple relation between $Q^{(m)}$ and $R^{(m)}$ holds,
 \beq
 R^{(m)} = - Q^{(m) \dagger}.
 \label{R_Q}
 \eeq
 It should be noted that the Hamiltonian and the Poisson
 bracket for the matrix mKdV equation (\ref{mmKdV1}) become
 invalid for the cmKdV equations (\ref{CMKdV5}) with $M \ge 3$. This is
 because the degree of freedom of $Q^{(m)}$ and $R^{(m)}$ for
 the cmKdV equations is less than that for the matrix mKdV equation. 

 \section{Inverse Scattering Method}
 \label{ISM}

 In this section we consider the scattering problem associated with 
 $2n \times 2n$ matrix 
(\ref{U_form}) under the constraint (\ref{R_eq_Q}) with $\ep = -1$ 
and the boundary conditions (\ref{boundary}), that is,
 \beq
\Psi_x = U \Psi, \hspace{4mm}
 U =
 \left[
 \begin{array}{cc}
  -\i \z I & Q \\
  R &  \i \z I\\
 \end{array}
 \right], \hspace{4mm} R=-Q^{\dagger},
 \label{scattering_problem2}
 \eeq
 \beq
 Q, R(=-Q^{\dagger}) \to O \hspace{5mm} {\rm as}
 \hspace{5mm} x \to \pm \infty.
 \label{boundary2}
 \eeq
 The results of this
 section are applicable to the matrix mKdV equation, the matrix NLS
 equation and other members of the hierarchy only if
 $Q$ and $R$ are $n \times n$ square matrices. 
The main idea in what follows is a modification of the analysis 
in refs.\ \citen{Wadati3}, \citen{Wadati4} for the matrix KdV 
equation. 

 \subsection{Scattering problem}

 Let $\Psi(\z)$ and $\Phi(\z)$ be 
solutions of eq.\ (\ref{scattering_problem2}) 
 composed by $2n$ rows and $n$ columns. We can show that 
 \beq
 \frac{\d }{\d x} \{ \Psi^{\dagger}(\z^{\ast}) \Phi(\z) \}= O.
 \label{}
 \eeq
 Hence we employ the following definition of $x$-independent 
 matrix Wronskian $W[\Psi, \Phi]$,
 \beq
 W[\Psi, \Phi] \equiv \Psi^{\dagger}(\z^{\ast}) \Phi(\z).
 \label{}
 \eeq

 We introduce Jost functions $\phi$, $\5{\phi}$ and  $\psi$, $\5{\psi}$ 
 which satisfy the boundary conditions,
\bseq
 \beq
\, \phi \sim 
 \left[
 \begin{array}{c}
   I  \\
   O  \\
 \end{array}
 \right]
 \e^{-\i \z x}
 \hspace{7mm}
 {\rm as}~~~ x \rightarrow -\infty,
 \label{phi}
 \eeq
 \beq
 \bar{\phi} \sim 
 \left[
 \begin{array}{c}
   O  \\
   -I  \\
 \end{array}
 \right]
 \e^{ \i \z x  \hphantom{-}}
 \hspace{5.4mm}
 {\rm as}~~~ x \rightarrow -\infty,
 \label{phi_bar}
 \eeq
 and
 \beq
 \psi \sim 
 \left[
 \begin{array}{c}
   O  \\
   I  \\
 \end{array}
 \right]
 \e^{ \i \z x \hphantom{-}}
 \hspace{7mm}
 {\rm as}~~~ x \rightarrow +\infty,
 \label{psi}
 \eeq
 \beq
 \bar{\psi} \sim 
 \left[
 \begin{array}{c}
   I  \\
   O  \\
 \end{array}
 \right]
 \e^{ - \i \z x }
 \hspace{6.9mm}
 {\rm as}~~~ x \rightarrow +\infty. \,
 \label{psi_bar}
 \eeq
\eseq
 Here $O$ and $I$ are respectively the $n \times n$ 
zero matrix and the $n \times n$ unit matrix. 
 It can be shown that $\phi \e^{\i \z x}$, $\psi \e^{-\i \z x}$
 are analytic in the upper 
 half plane of $\z$, and $\5{\phi} \e^{-\i \z x}$, 
$\5{\psi}\e^{\i \z x}$ are analytic 
 in the lower half plane of $\z$ because $Q$, $R$ are assumed to
 go to $O$ sufficiently rapidly at $x \to \pm \infty$. 
 We assume the following integral representation of the Jost functions 
 $\psi$ and $\5{\psi}$,
 \beq
 \psi =
 \left[
 \begin{array}{c}
   O  \\
   I  \\
 \end{array}
 \right]
 \e^{ \i \z x}
 +\int_{x}^{\infty}K(x,s)\e^{\i \z s} \d s,
 \label{psi_form}
 \eeq
 \beq
 \bar{\psi} =
 \left[
 \begin{array}{c}
   I  \\
   O  \\
 \end{array}
 \right]
 \e^{ -\i \z x}
 +\int_{x}^{\infty} \bar{K}(x,s)\e^{- \i \z s} \d s,
 \label{psi_bar_form}
 \eeq
 where $K(x,s)$ and $\5{K}(x,s)$ are column vectors
  whose elements are $n \times n$ square matrices,
 \beq
 K(x,s) =
 \left[
 \begin{array}{c}
  K_1(x,s)  \\
  K_2(x,s)  \\
 \end{array}
 \right],
 \bar{K}(x,s) =
 \left[
 \begin{array}{c}
  \bar{K}_1(x,s)  \\
  \bar{K}_2(x,s)  \\
 \end{array}
 \right].
 \label{}
 \eeq
 We substitute eq.\ (\ref{psi_form}) for eq.\ (\ref{scattering_problem2}) 
 and get the relations for $K_1$ and $K_2$,
 \beq
 \lim_{s \rightarrow +\infty}\left[
 \begin{array}{c}
  K_1(x,s)  \\
  K_2(x,s)  \\
 \end{array}
 \right]
 =
 \left[
 \begin{array}{c}
  O  \\
  O  \\
 \end{array}
 \right],
 \label{}
 \eeq
 \beq
 -2K_1(x,x) = Q(x),
 \label{}
 \eeq
 \beq
 (\6_x - \6_s) K_1(x,s) = Q(x) K_2(x,s) \hspace{6mm}  (s>x),
 \label{}
 \eeq
 \beq
 (\6_x + \6_s) K_2(x,s) = R(x) K_1(x,s) \hspace{6mm}  (s>x).
 \label{}
 \eeq
 Similarly, substituting eq.\ (\ref{psi_bar_form}) for
 eq.\ (\ref{scattering_problem2}), we get
 for $\5{K}_1$ and $\5{K}_2$,
 \beq
 \lim_{s \rightarrow +\infty}\left[
 \begin{array}{c}
  \bar{K}_1(x,s)  \\
  \bar{K}_2(x,s)  \\
 \end{array}
 \right]
 =
 \left[
 \begin{array}{c}
  O  \\
  O  \\
 \end{array}
 \right],
 \label{}
 \eeq
 \beq
 -2\bar{K}_2(x,x) = R(x),
 \label{}
 \eeq
 \beq
 (\6_x - \6_s) \bar{K}_2(x,s) = R(x) \bar{K}_1(x,s) \hspace{6mm}  (s>x),
 \label{}
 \eeq
 \beq
 (\6_x + \6_s) \bar{K}_1(x,s) = Q(x) \bar{K}_2(x,s) \hspace{6mm}  (s>x).
 \label{}
 \eeq

 Because a pair of the Jost functions $\phi$ and $\5{\phi}$, or 
 $\psi$ and $\5{\psi}$ forms a fundamental system of solutions of 
 eq.\ (\ref{scattering_problem2}), we can set
\bseq
 \beq
 \phi(x,\z) = \bar{\psi}(x,\z)A(\z) + \psi(x,\z)B(\z),
 \label{phi_relation}
 \eeq
 \beq
 \bar{\phi}(x,\z) = \bar{\psi}(x,\z) \bar{B}(\z) - \psi(x,\z) \bar{A}(\z).
 \label{phi_bar_relation}
 \eeq
\label{phi_5_phi}
\eseq
 Here the coefficients $A(\z)$, $\5{A}(\z)$, $B(\z)$ and $\5{B}(\z)$ 
 are $x$-independent $n \times n$ matrices and called scattering data. 

 According to the asymptotic behaviors of the Jost functions 
 (\ref{phi})--(\ref{psi_bar}), we get
\bseq
 \bea
    W[ \phi, \phi] = W[ \5{\phi}, \5{\phi}]
  &=& W[ \psi, \psi] = W[ \5{\psi}, \5{\psi}]
  = I,
 \label{w1} \\
% \eeq
 %
% \beq
    W[\5{\phi}, \phi] &=& W[ \5{\psi}, \psi]
  = O,
 \label{w2} \\
% \eeq
 %
 %
% \beq
 A(\z) &=& W[ \5{\psi}, \phi],
 \label{rep1} \\
% \eeq
 %
% \beq
 \5{A}(\z) &=& - W[ \psi, \5{\phi}],
 \label{rep2} \\
% \eeq
% \beq
 B(\z) &=& W[ \psi, \phi],
 \label{rep3} \\
% \eeq
% \beq
 \5{B}(\z) &=& W[ \5{\psi}, \5{\phi}].
\label{rep4}
\eea
\label{}
\eseq
 The expressions (\ref{rep1})--(\ref{rep2}) show
  that $A(\z)$ and $\5{A}(\z)$ are analytic 
 respectively in the upper half plane and in the lower half plane. 
 Using the above relations (\ref{w1})--(\ref{rep4}), 
 we obtain the following relations among 
$A(\z)$, $\5{A}(\z)$, $B(\z)$ and $\5{B}(\z)$,
\bseq
 \beq
 A^{\dagger}(\z^{\ast})A(\z) + B^{\dagger}(\z^{\ast})B(\z) = I,
 \label{scat_rel1}
 \eeq
 \beq
 \5{A}^{\dagger}(\z^{\ast})\5{A}(\z) +
  \5{B}^{\dagger}(\z^{\ast})\5{B}(\z) = I,
 \label{scat_rel2}
 \eeq
 \beq
 A^{\dagger}(\z^{\ast})\5{B}(\z) - B^{\dagger}(\z^{\ast})\5{A}(\z) = O.
 \label{scat_rel3}
 \eeq
\eseq
These relations are written as
\beq
\left[
\begin{array}{cc}
 A^\dagger(\z^\ast) & B^\dagger(\z^\ast) \\
 \5{B}^\dagger(\z^\ast) & -\5{A}^\dagger(\z^\ast) \\
\end{array}
\right]
\left[
\begin{array}{cc}
 A(\z) & \5{B}(\z) \\
 B(\z) & -\5{A}(\z) \\
\end{array}
\right]
=
\left[
\begin{array}{cc}
 I & O \\
 O & I \\
\end{array}
\right],
\label{}
\eeq
which leads to the inversion of eq.\ (\ref{phi_5_phi}),
\bseq
 \beq
 \5{\psi}(x,\z) = \phi(x,\z)A^\dagger(\z^\ast) 
+ \5{\phi}(x,\z)\5{B}^\dagger(\z^\ast),
 \label{psi_relation}
 \eeq
 \beq
 \psi(x,\z) = \phi(x,\z) B^\dagger(\z^\ast) 
- \5{\phi}(x,\z) \bar{A}^\dagger(\z^\ast).
 \label{psi_bar_relation}
 \eeq
\label{psi_5_psi}
\eseq%

 \subsection{Gel'fand-Levitan-Marchenko equation}
 \label{}
 To derive the formula of the ISM concisely, we assume that $A(\z)$, 
 $\5{A}(\z)$, $B(\z)$ and $\5{B}(\z)$ are entire functions. This
 assumption is true if the potentials $Q$ and $R$ decrease faster than 
 any exponential function at $x \to \pm \infty$. The result is, however, 
 applicable to larger classes of potentials $Q$ and $R$. 

 Multiplying $A(\z)^{-1}$ and $\5{A}(\z)^{-1}$ from the right to eqs.\ 
 (\ref{phi_relation}) and (\ref{phi_bar_relation}) respectively, we get
\bseq
 \beq
 \phi(x,\z) A(\z)^{-1} = \bar{\psi}(x,\z) + \psi(x,\z)B(\z)A(\z)^{-1},
 \label{prep1}
 \eeq
 \beq
 \bar{\phi}(x,\z) \bar{A}(\z)^{-1} 
 = - \psi(x,\z) + \bar{\psi}(x,\z) \bar{B}(\z) \bar{A}(\z)^{-1}.
 \label{prep2}
 \eeq
\eseq
We operate
 \beq
 \frac{1}{2\pi}\int_C \d \z \e^{\i \z y} \hspace{5mm} (y>x)
 \label{}
 \eeq 
to eq.\ (\ref{prep1}), where $C$ is a semi-circle contour from
$-\infty+\i 0^+$ to $+\infty + \i 0^+$ passing above all poles of $\{
{\rm det} A(\z)\}^{-1}$. After a standard calculation, we get the Gel'fand-Levitan-Marchenko equation,
 \beq
 \bar{K}(x,y) + 
 \left[
 \begin{array}{c}
   O  \\
   I  \\
 \end{array}
 \right] F(x+y)
 + \int_x^{\infty} K(x,s) F(s+y) \d s = 
 \left[
 \begin{array}{c}
  O  \\
  O  \\
 \end{array}
 \right]
 \hspace{10mm} (y>x),
 \label{GL1}
 \eeq
 where $F(x)$ is defined by
 \beq
 F(x) = \frac{1}{2\pi} \int_{C} \d \z \e^{\i \z x}B(\z)A(\z)^{-1}.
 \label{}
 \eeq
 We remark that $A(\z)^{-1}$ is given by
 \beq
 A(\z)^{-1} = \frac{1}{\det A(\z)} \tilde{A}(\z),
 \label{}
 \eeq
 where $\tilde{A}$ is the cofactor matrix of $A$. 
We assume that $1/\det A(\z)$ has $N$ 
isolated simple poles $\{\z_1, \z_2, \cdots, \z_N \}$ in the upper
half plane and is
 regular on the real axis. Each of these poles
 determines one bound state. Then,
  by the use of the residue theorem, we get an alternative expression of $F$,
 \beq
 F(x) = \frac{1}{2\pi} \int_{-\infty}^{\infty}
        \d \xi \e^{\i \xi x}B(\xi)A(\xi)^{-1}
        - \i \sum_{j=1}^N C_j \e^{\i \z_j x}.
 \label{F_form}
 \eeq
 Here $C_j$ is the residue matrix of $B(\z) A(\z)^{-1}$ at $\z = \z_j$.

Similarly, we operate
 \beq
 \frac{1}{2\pi}\int_{\bar{C}} \d \z \e^{-\i \z y} \hspace{5mm} (y>x)
 \label{}
 \eeq 
 to eq.\ (\ref{prep2}), where $\bar{C}$ is a semi-circle contour
  from $-\infty+\i 0^-$
 to $+\infty + \i 0^-$ passing below all poles of $\{{\rm det}\bar{A}(\z)\}
 ^{-1}$. We get the counterpart of the Gel'fand-Levitan-Marchenko 
 equation,
 \beq
 K(x,y) -
 \left[
 \begin{array}{c}
   I  \\
   O  \\
 \end{array}
 \right] \bar{F}(x+y)
 - \int_x^{\infty} \bar{K}(x,s) \bar{F}(s+y) \d s = 
 \left[
 \begin{array}{c}
  O  \\
  O  \\
 \end{array}
 \right]
 \hspace{10mm} (y>x),
 \label{GL2}
 \eeq
 where $\bar{F}(x)$ is defined by
 \beq
 \bar{F}(x) = \frac{1}{2\pi} \int_{\bar{C}} \d \z \e^{-\i \z x}\bar{B}
 (\z)\bar{A}(\z)^{-1}.
 \label{}
 \eeq
If we assume that $1/\det \5{A}(\z)$ has $\5{N}$ isolated simple
 poles $\{\5{\z}_1, \5{\z}_2, \cdots, \5{\z}_{\5{N}} \}$ in the lower 
half plane and is regular on the real axis, we get an alternative
 expression of $\5{F}$,
 \beq
 \bar{F}(x) = \frac{1}{2\pi} \int_{-\infty}^{\infty}
        \d \xi \e^{-\i \xi x}\bar{B}(\xi) \bar{A}(\xi)^{-1}
        + \i \sum_{k=1}^{\bar{N}} \bar{C}_k \e^{- \i \bar{\z}_k x}.
 \label{F_bar_form}
 \eeq
 Here $\5{C}_k$ is the residue matrix of $\5{B}(\z) \5{A}(\z)^{-1}$
  at $\z = \5{\z}_k$.

 \subsection{Time-dependence of the scattering data}
 \label{}

 Under the rapidly decreasing boundary conditions (\ref{boundary2}), 
 the asymptotic form of the Lax matrix $V$ is given by
 \beq
 V \rightarrow 
 \left[
 \begin{array}{cc}
  -4\i \z^3 I  &  O \\
   O  &  4\i \z^3 I \\
 \end{array}
 \right]
 \hspace{5mm} {\rm as} \hspace{2mm}x \rightarrow \pm \infty.
 \label{}
 \eeq

 We define time-dependent Jost functions by
\bseq
 \bea
 \phi^{(t)} &\equiv& \phi \e^{-4\i \z^3 t} \sim
 \left[
 \begin{array}{c}
  I \\
  O \\
 \end{array}
 \right]\vphantom{\sum_{\stackrel{B}{<}}}
 \e^{-\i \z x - 4\i \z^3 t}
 \hspace{7mm}
 {\rm as}~~~ x \rightarrow -\infty,
\\ 
% \eeq
 %
 %
% \beq
 \bar{\phi}^{(t)} &\equiv& \bar{\phi} \e^{4\i \z^3 t} \sim
 \left[
 \begin{array}{c}
  O \\
  -I \\
 \end{array}
 \right]\vphantom{\sum_{\stackrel{B}{<}}}
 \e^{\i \z x + 4\i \z^3 t}
 \hspace{9.7mm}
 {\rm as}~~~ x \rightarrow -\infty,
\\ 
% \eeq
 %
 %
% \beq
 \psi^{(t)} &\equiv& \psi \e^{4\i \z^3 t} \sim
 \left[
 \begin{array}{c}
  O \\
  I \\
 \end{array}
 \right]\vphantom{\sum_{\stackrel{B}{<}}}
 \e^{\i \z x + 4\i \z^3 t}
 \hspace{11.6mm}
 {\rm as}~~~ x \rightarrow +\infty,
\\
% \eeq
 %
 %
% \beq
 \bar{\psi}^{(t)} &\equiv& \bar{\psi} \e^{-4\i \z^3 t} \sim
 \left[
 \begin{array}{c}
  I \\
  O \\
 \end{array}
 \right]
 \e^{-\i \z x - 4\i \z^3 t}
 \hspace{6.9mm}
 {\rm as}~~~ x \rightarrow +\infty.
\eea
\eseq

 From the relations
 \beq
 \frac{\6 \phi^{(t)}}{\6 t} = V \phi^{(t)}, \hspace{5mm}
 \frac{\6 \bar{\phi}^{(t)}}{\6 t} = V \bar{\phi}^{(t)}, 
 \label{}
 \eeq
 we get
 \beq
 \frac{\6 \phi}{\6 t} = (V + 4\i \z^3 I) \phi,
 \hspace{5mm} \frac{\6 \bar{\phi}}{\6 t} = (V - 4\i \z^3 I) \bar{\phi}.
 \label{time1}
 \eeq

 We substitute the definitions of the scattering data,
\bseq
 \beq
 \phi(x,\z) = \bar{\psi}(x,\z)A(\z, t) + \psi(x,\z)B(\z, t),
 \label{time3}
 \eeq
 \beq
 \bar{\phi}(x,\z) = \bar{\psi}(x,\z) \bar{B}(\z, t) - \psi(x,\z)
 \bar{A}(\z, t),
 \label{time4}
 \eeq
\label{time_total}
\eseq
into eq.\ (\ref{time1}). Then taking the limit 
 $x \rightarrow +\infty$, we get
\bseq
 \beq
 A_t (\z, t) = O,
 \label{}
 \eeq
 \beq
 B_t (\z, t) = 8 \i \z^3 B(\z, t),
 \label{}
 \eeq
 and
 \beq
 \bar{A}_t (\z, t) = O,
 \label{}
 \eeq
 \beq
 \bar{B}_t (\z, t) = -8 \i \z^3 \bar{B}(\z, t).
 \label{}
 \eeq
\eseq
 The above relations lead to the following time-dependence of 
 the scattering data:
\bseq
 \beq
 A(\z, t) = A(\z, 0),
 \label{A_time}
 \eeq
 \beq
 B(\z, t) = B(\z, 0)\e^{8\i \z^3 t},
 \label{B_time}
 \eeq
 and
 \beq
 \bar{A}(\z, t) = \bar{A}(\z, 0),
 \label{A_bar_time}
 \eeq
 \beq 
 \bar{B}(\z, t) = \bar{B}(\z, 0) \e^{-8\i \z^3 t}.
 \label{B_bar_time}
 \eeq
\eseq
By eqs.\ (\ref{A_time})--(\ref{B_bar_time}),
 the time-dependences of $BA^{-1}$, $C_j$ and $\5{B} \5{A}^{-1}$,
 $\bar{C}_k$ are respectively given by
 \beq
 B(\xi,t) A(\xi, t)^{-1} = B(\xi, 0)A(\xi, 0)^{-1} e^{8 \i \xi^3 t},
 \label{BA_time}
 \eeq
 \beq
 C_j(t) = C_j(0)e^{8 \i \z_j^3 t},
 \label{C_time}
 \eeq
 and
 \beq
 \5{B}(\xi,t) \5{A}(\xi, t)^{-1} 
 = \5{B}(\xi, 0) \5{A}(\xi, 0)^{-1} e^{-8 \i \xi^3 t}.
 \label{BA_bar_time}
 \eeq
 \beq
 \bar{C}_k(t) = \bar{C}_k(0)e^{-8 \i \bar{\z}_k^3 t}. 
 \label{C_bar_time}
 \eeq
To summarize, we obtain explicit time-dependent 
forms of $F(x, t)$ and $\5{F}(x, t)$,
 \beq
 F(x, t) = \frac{1}{2\pi} \int_{-\infty}^{\infty}
        \d \xi \e^{\i \xi x +8\i \xi^3 t} B(\xi,0)A(\xi,0)^{-1}
        - \i \sum_{j=1}^N C_j(0) \e^{\i \z_j x +8\i \z_j^3 t},
 \label{ref55}
 \eeq
 \beq
 \5{F}(x,t) = \frac{1}{2\pi} \int_{-\infty}^{\infty}
        \d \xi \e^{- \i \xi x - 8\i \xi^3 t} \5{B}(\xi,0)\5{A}(\xi,0)^{-1}
        + \i \sum_{k=1}^{\5{N}} \5{C}_k(0) \e^{-\i \5{\z}_k x - 8\i \5{\z}_k^3 t}.
 \label{}
 \eeq
% where $B(\z) A(\z)^{-1}$, $\5{B}(\z)\5{A}(\z)^{-1}$, $C_j$, $\5{C}_k$
% are time-independent matrices. 
%
 \subsection{Initial value problem}
 \label{Initial value problem}
Because of the constraint $R=-Q^{\dagger}$, we have some relations
 which make the further analysis simple.

 First, we have
 \beq
 \det \5{A}(\z) = \{\det A (\z^{\ast})\}^{\ast},
 \label{det_A2}
 \eeq
 which is proved in Appendix A. This relation gives us a useful 
information
  about the total number and
 the positions of the poles of $A(\z)^{-1}$ and $\5{A}(\z)^{-1}$,
 \beq
 \5{N} = N, \hspace{5mm} \5{\z}_k= \z_k^{\ast}.
 \label{eq_ref11}
 \eeq

 Second, due to eq.\ (\ref{scat_rel3}), we have
 \beq
 \5{B}(\z)\5{A}(\z)^{-1} = \{ B(\z^{\ast}) A(\z^{\ast})^{-1}\}
 ^{\dagger},
 \label{}
 \eeq
 which leads to 
 \bseq
 \beq
 \5{B}(\xi)\5{A}(\xi)^{-1} = \{ B(\xi) A(\xi)^{-1}\}^{\dagger}
 \hspace{2mm} (\xi : {\rm real}),
 \label{}
 \eeq
 \beq
 \5{C}_k = C_k^{\dagger}. \hspace{15mm}
 \label{}
 \eeq
 \label{scat_rel4}
 \eseq
The relations (\ref{eq_ref11}) and 
(\ref{scat_rel4}) give a connection between $\5{F}(x, t)$ and $F(x, t)$,
 \beq
 \5{F}(x,t) = F(x,t)^{\dagger}.
 \label{F_bar_F}
 \eeq
Combining the above results, we arrive at
 \beq
 K_1 (x,y ;t) = F(x+y ,t)^{\dagger} - \int_x^{\infty} 
 \d s_1 \int_x ^{\infty} \d s_2 K_1 (x, s_2;t) F(s_2 + s_1,t) 
 F(s_1 +y,t)^{\dagger},
 \label{GL3}
 \eeq
 \beq
 \5{K}_2 (x,y;t) = -F(x+y,t) - \int_x^{\infty} 
 \d s_1 \int_x ^{\infty} \d s_2 \5{K}_2 (x, s_2;t) 
 F(s_2 + s_1,t)^{\dagger} F(s_1 +y,t),
 \label{GL4}
 \eeq
where $F(x,t)$ is given by eq. (\ref{ref55}). 
% \beq
% F(x;t) = \frac{1}{2\pi} \int_{-\infty}^{\infty}
%        \d \xi \e^{\i \xi x}B(\xi,t)A(\xi,t)^{-1}
%        - \i \sum_{j=1}^N C_j(t) \e^{\i \z_j x}.
% \label{F_form2}
% \eeq
%

We can solve the initial value problem of the matrix mKdV equation 
 as follows. \\
 (1) For given potentials at $t=0$, $Q(x, 0)$ and $R(x, 0)$ which satisfy 
 $R(x,0)=-Q(x,0)^{\dagger}$, we solve the 
 scattering problem (\ref{scattering_problem2}), and obtain 
 scattering data $ \{ B(\xi)A(\xi)^{-1}, \z_j, \i C_j \}$.
\\
 (2) The time-dependence of the scattering data is determined by 
 eqs.\ (\ref{BA_time}) and (\ref{C_time}).
 \\
 (3) We substitute the time-dependent scattering data into
 the Gel'fand-Levitan-Marchenko equations (\ref{GL3}) and (\ref{GL4}). 
 Solving the equations, we reconstruct the time-dependent potentials,
 \beq
 Q(x, t) = -2K_1(x,x;t),
 \label{}
 \eeq
 \beq
 R(x, t) = -2\5{K}_2(x,x;t).
 \label{}
 \eeq
In this way, we obtain the solution $Q(x,t)$ and $R(x,t)$ from 
the initial condition $Q(x,0)$ and $R(x,0)$.
 This procedure proves directly the complete integrability of the matrix 
 mKdV equation (\ref{mmKdV2}) with $\ep = -1$. 

 If we employ other time-dependences of the scattering data, for
instance, 
 \beq
 B(\xi,t) A(\xi, t)^{-1} = B(\xi, 0)A(\xi, 0)^{-1} e^{4 \i \xi^2 t},
 \label{BA_time2}
 \eeq
 \beq
 C_j(t) = C_j(0)e^{4 \i \z_j^2 t},
 \label{C_time2}
 \eeq
 the initial value problem of the matrix NLS equation 
 (\ref{mNLS2}) with $\ep = -1$ can be solved. 

 As for the constraint $R=-Q^{\dagger}$, one comment is in order. 
 Because $\5{F}$ is connected with $F$
  by eq.\ (\ref{F_bar_F}), we can prove by the Neumann-Liouville
 expansion (see Appendix B) that the
 solution of eqs.\ (\ref{GL3}) and (\ref{GL4}) satisfies 
 \beq
 \5{K}_2(x,x;t)= -K_1(x,x;t)^{\dagger}.
 \label{}
 \eeq
 This relation assures that the relation $R(x,t)=-Q(x,t)^{\dagger}$ 
 holds at any time $t$.

 \subsection{Soliton solutions}
 \label{}
 Assuming the reflection-free potentials which satisfy
 \beq
 B(\xi)=\5{B}(\xi)= O \hspace{5mm}(\xi : {\rm real}),
 \label{}
 \eeq
 we can construct soliton solutions of the matrix mKdV equation. 
 In this case $F(x)$ is given by
 \beq
 F(x,t) = - \i \sum_{j=1}^N C_j(t) \e^{\i \z_j x},
 \label{F_ref_less}
\eeq
\beq
 C_j(t)= C_j(0) \e^{8\i \z_j^3 t}. \hspace{7.0mm}
\label{C_mmKdV}
\eeq
To solve eq.\ (\ref{GL3}) with eq.\ (\ref{F_ref_less}), we set
 \beq
 K_1 (x, y;t) = \i \sum_{k=1}^N P_k (x,t) C_k(t)^{\dagger}
 \e^{-\i \z_k^\ast (x+y)}.
 \label{K_1_form}
 \eeq
 Introducing eq.\ (\ref{K_1_form}) into eq.\ (\ref{GL3}), we have 
a set of algebraic equations,
 \beq
 P_k (x,t)- \sum_{l=1}^N \sum_{j=1}^N 
 \frac{1}{(\z_j -\z_k^{\ast})(\z_j-\z_l^{\ast})} P_l (x,t)
 C_l(t)^{\dagger} C_j (t) \e^{2\i (\z_j -\z_l^{\ast})x} = I.
 \label{P_eq}
 \eeq
 We define a matrix $S$ by
 \beq
 S_{lk} \equiv \de_{lk}I -\sum_{j=1}^N 
 \frac{\e^{2\i (\z_j -\z_l^{\ast})x}}
 {(\z_j -\z_k^{\ast})(\z_j-\z_l^{\ast})} C_l(t)^\dagger C_j(t),
 \hspace{5mm} 1 \le l, k \le N.
 \label{}
 \eeq
 Then eq.\ (\ref{P_eq}) is expressed as
 \beq
 (\; P_1 \; P_2 \; \cdots \; P_N \; )
 \left(
 \begin{array}{ccc}
  S_{11} & \cdots & S_{1N}\\
  \vdots & \ddots & \vdots\\
  S_{N1} & \cdots & S_{NN} \\
 \end{array}
 \right)
 =
 (\, \underbrace{\, I \; I \; \cdots \; I \,}_{N} \, ).
 \label{}
 \eeq
 Thus the $N$-soliton solution of the matrix mKdV equation
 (\ref{mmKdV2}) with $\ep=-1$ is
  given by 
 \bea
 Q(x,t) &=& -2 K_1(x,x;t)
 \nn \\
 &=& -2 \i \sum_{k=1}^N P_k (x,t) C_k(t)^{\dagger} 
  \e^{-2 \i \z_k^\ast x}
 \nn \\
 &=& -2 \i \, (\, \underbrace{\, I \; I \; \cdots \; I \,}_{N} \, )
 \; S^{-1} \;
 \left(
 \begin{array}{c}
  C_1 (t)^{\dagger} \e^{-2\i \z_1^\ast x}   \\
  C_2 (t)^{\dagger} \e^{-2\i \z_2^\ast x}   \\
    \vdots  \\
  C_N (t)^{\dagger} \e^{-2\i \z_N^\ast x}   \\
 \end{array}
 \right).
 \label{N-soliton}
 \eea
 As a special case $N=1$, we get 1-soliton solution of the 
matrix mKdV equation,
 \bea
 Q(x,t) 
 &=& 
  -2 \i \biggl\{ I
 -\frac{\e^{8 \i (\z_1^3 - \z_1^{\ast\,3})t}}{(\z_1 -\z_1^{\ast})^2}
 C_1(0)^{\dagger}C_1(0) \e^{2\i (\z_1-\z_1^{\ast})x}\biggr\}^{-1}
 C_1(0)^{\dagger} \e^{-2\i \z_1^{\ast} x -8 \i \z_1^{\ast \, 3} t}
 \nn \\
 &=& 
  -2 \i 
 \biggl\{ \e^{-\i(\z_1-\z_1^{\ast})x-4\i(\z_1^3-\z_1^{\ast \, 3})t} I
 -\frac{1}{(\z_1 -\z_1^{\ast})^2}
 C_1(0)^{\dagger}C_1(0) \e^{\i (\z_1-\z_1^{\ast})x+4\i
 (\z_1^3-\z_1^{\ast \, 3})t} \biggr\}^{-1}
 \nn \\
 && \hspace{4mm} 
 \cdot C_1(0)^{\dagger} \e^{-\i (\z_1+\z_1^{\ast}) x
  -4 \i (\z_1^3+\z_1^{\ast \, 3}) t}.
 \label{}
 \eea
If we replace the time dependence (\ref{C_mmKdV})
 in eq.\ (\ref{F_ref_less}) with eq.\ (\ref{C_time2}),
 we obtain the 
$N$-soliton solution of the matrix NLS equation (\ref{mNLS2}) 
with $\ep=-1$.

 \section{Reduction of the ISM for the Coupled Modified KdV Equations}
 \label{Reduction of the ISM for the cmKdV equation}

 In order to make the ISM in \S \ref{ISM} applicable to the cmKdV
 equations, we have to take into account the internal symmetry 
 of $Q$ and $R$ defined by eqs.\ (\ref{QR_def4})--(\ref{QR_def6}).

 If we set $Q^{(m)}$ and $R^{(m)}$ for $m\ge 2$ as
 \beq
 Q^{(m)}= v_0 \eins 
%{\bf 1}
+ \sum_{k=1}^{2m-1} v_k e_k, 
 \hspace{5mm}
 R^{(m)}= -v_0 \eins 
%{\bf 1}
+ \sum_{k=1}^{2m-1} v_k e_k, 
 \label{}
 \eeq
 the following important 
relations for $2^{m-1}\times 2^{m-1}$ matrices 
$\{e_i\}$ hold,
 \bea
 \{ e_i, e_j \}_+ &=& -2 \de_{ij} \eins, 
 \label{ei_ej}
 \\
 e_k^{\dagger} &=& - e_k, \hspace{8.5mm}
 \\
 \tr \, e_k &=& 0. \hspace{8.5mm}
 \label{tr_ei}
 \eea
Here $\{\cdot, \cdot \}_+$ denotes the anti-commutator. $\eins$ is 
the $2^{m-1}\times 2^{m-1}$ unit matrix. Equation (\ref{ei_ej}) 
leads to 
\beq
\tr (e_i e_j) = -2^{m-1} \de_{ij}. 
\label{tr_ei_ej}
\eeq

The results in 
 \S \ref{Conservation laws} assures that the cmKdV equations
 have an infinite number of conservation laws.  From the explicit 
forms of the conserved quantities, we find that the first four
 conserved densities for the original cmKdV equations 
(\ref{CMKdV}) are given by 
\beq
\sum_{j,k} C_{jk} u_j u_k,
\label{}
\eeq
\beq
u_j u_{k,x}, \hspace{3mm} \forall \, j,\hspace{0.5mm} k 
\hspace{2mm} (j \neq k),
\label{quantity_special}
\eeq
\beq
\biggl( \sum_{j,k} C_{jk} u_j u_k \biggr)^2 -
 \sum_{j,k} C_{jk} u_{j,x} u_{k,x},
\label{}
\eeq
\beq
\biggl( \sum_{j,k} C_{jk} u_j u_k \biggr)^3 -
 3 \sum_{j,k} C_{jk} u_{j} u_{k} \cdot
 \sum_{j,k} C_{jk} u_{j,x} u_{k,x}
+\frac{1}{2} \sum_{j,k} C_{jk} u_{j,xx} u_{k,xx}
-\frac{1}{2} \biggl\{ 
\biggl(\sum_{j,k} C_{jk} u_{j} u_{k}\biggr)_x \biggr\}^2.
\label{}
\eeq
We remark that the method in \S \ref{Conservation laws} does not 
give the quantity (\ref{quantity_special}). 

Next, we discuss the initial value problem and the soliton solutions 
of the cmKdV equations. 
Considering the scattering problem (\ref{scattering_problem2}) 
with the potentials $Q^{(m)}$ and $R^{(m)}$ for $m \ge 2$, we 
can show that there are following 
restrictions on the scattering data. \\
{\bf Proposition 4.1} \\
(1) The determinant of $A(\z)$ satisfies
\beq
\det {A}(\z) = \{\det A (- \z^{\ast})\}^{\ast},
\label{}
\eeq
as a function of complex $\z$. 
Thus the poles of $1/\det A(\z)$ in the upper half plane 
should appear on the imaginary axis or as pairs which are 
situated symmetric 
with respect to the imaginary axis. Therefore, 
% we replace $N$ in \$ 3 with $2N$ and 
we can set the values 
of $2N$ poles as 
\bseq
\bea
&& \z_{2j-1} = \xi_j + \i \eta_j, \hspace{5mm} j=1, 2, \cdots, N,
\label{}
\\
%\eeq
%
%\beq
&& \z_{2j} = - \z_{2j-1}^{\ast} = -\xi_j + \i \eta_j,
\hspace{5mm} j=1, 2, \cdots, N,
\label{}
\eea
\label{z_z}
\eseq
for $\eta_j > 0$. The condition (\ref{z_z}) should be 
interpreted as follows; if $\z_i$ is pure imaginary, 
it does not need its counterpart. 
\\
(2) The reflection coefficient 
$B(\xi) A(\xi)^{-1}$ for real $\xi$ should be expressed as
\beq
B(\xi) A(\xi)^{-1} = r^{(0)} \eins + \sum_{k=1}^{2m-1}r^{(k)} e_k.
\label{}
\eeq
Here $r^{(0)}$ and $r^{(k)}$ are complex functions of $\xi$, $t$ 
which satisfy
\beq
r^{(0)}(-\xi) = r^{(0)}(\xi)^{\ast}, \hspace{2mm}
r^{(k)}(-\xi) = r^{(k)}(\xi)^{\ast}.
\label{r_r}
\eeq
(3) The residue matrices $\{ C_1, C_2, \cdots, C_{2N-1}, C_{2N}\}$ 
should be expressed as
\bseq
\beq
\i C_{2j-1} = c_j^{(0)} \eins + \sum_{k=1}^{2m-1} c_j^{(k)} e_k,
\hspace{5mm} j=1, 2, \cdots, N,
\label{}
\eeq
\beq
\i C_{2j} = c_j^{(0)\, \ast} \eins 
+ \sum_{k=1}^{2m-1} c_j^{(k)\, \ast} e_k,
\hspace{5mm} j=1, 2, \cdots, N,
\eeq
\label{C_C}
\eseq
where $c_j^{(0)}$, $c_j^{(k)}$ are complex constants. 
For example, $\{ C_1, C_2, \cdots, C_{2N} \}$ for 
the 4-component cmKdV equations are given by 
\bseq
\beq
\i C_{2j-1} =
\left[
\begin{array}{cc}
 \a_j & \beta_j \\
 -\gamma_j & \de_j \\
\end{array}
\right],
\label{}
\eeq
\beq
\i C_{2j} =
\left[
\begin{array}{cc}
 \de_j^\ast & \gamma_j^\ast \\
 -\beta_j^\ast & \a_j^\ast \\
\end{array}
\right],
\label{}
\eeq
\eseq
in a different notation. A proof of the statements is given in Appendix C. 

Considering the above conditions, we have explicit 
expressions of $F$ and $\5{F}$ in terms of $\eins$ and $e_k$, 
\bea
 F(x, t) &=& \frac{1}{2\pi} \int_{-\infty}^{\infty}
        B(\xi,t) A(\xi,t)^{-1}
        \e^{\i \xi x} \d \xi 
         - \i \sum_{j=1}^{2N} C_j(t) \e^{\i \z_j x}
\\
&=&
\frac{1}{2\pi} \int_{0}^{\infty}
        \biggl\{ (r^{(0)}\e^{\i \xi x}+r^{(0)\, \ast}\e^{-\i \xi x}) \eins
         + \sum_{k=1}^{2m-1}
        ( r^{(k)} \e^{\i \xi x} + r^{(k)\, \ast}\e^{-\i \xi x}) e_k \biggr\}
        \, \d \xi 
\nn \\
&&       - \sum_{j=1}^{N} \biggl\{ (c_j^{(0)} \e^{\i \z_j x}
        +c_j^{(0)\, \ast} \e^{-\i \z_j^{\ast} x})\eins
         + \sum_{k=1}^{2m-1} (c_j^{(k)}\e^{\i \z_j x}
         +c_j^{(k)\, \ast}\e^{-\i \z_j^{\ast} x}) e_k \biggr\} ,
\label{F_form3}
\eea
\bea
 \5{F}(x, t) &=& F(x, t)^{\dagger} 
\nn \\
&=& 
\frac{1}{2\pi} \int_{-\infty}^{\infty}
%       \biggl(r_0 \eins - \sum_{k=1}^{2m-1} r_k e_k \biggr)
        \5{B}(\xi,t)\5{A}(\xi,t)^{-1}
        \e^{-\i \xi x} \d \xi 
         + \i \sum_{j=1}^{2N} C_j(t)^{\dagger} \e^{-\i \z_j^{\ast} x}
\\
&=&
\frac{1}{2\pi} \int_{0}^{\infty}
        \biggl\{ (r^{(0)}\e^{\i \xi x}+r^{(0)\, \ast}\e^{-\i \xi x}) \eins
         - \sum_{k=1}^{2m-1}
        ( r^{(k)} \e^{\i \xi x} + r^{(k)\, \ast}\e^{-\i \xi x})
 e_k \biggr\}
        \, \d \xi 
\nn \\
&&       - \sum_{j=1}^{N} \biggl\{ (c_j^{(0)} \e^{\i \z_j x}
        +c_j^{(0)\, \ast} \e^{-\i \z_j^{\ast} x}) \eins
%{\bf 1}
         - \sum_{k=1}^{2m-1} (c_j^{(k)}\e^{\i \z_j x}
         +c_j^{(k)\, \ast}\e^{-\i \z_j^{\ast} x}) e_k \biggr\} .
\label{F_bar_form3}
\eea
Because $B(\xi,t)A(\xi,t)^{-1}$ and $C_j(t)$ depend on $t$ as
 \beq
 B(\xi,t) A(\xi, t)^{-1} = B(\xi, 0)A(\xi, 0)^{-1} e^{8 \i \xi^3 t},
 \label{BA_time3}
 \eeq
 \beq
 C_j(t) = C_j(0)e^{8 \i \z_j^3 t},
 \label{C_time3}
 \eeq
the time-dependences of $r^{(0)}$, $r^{(k)}$
 and $c_j^{(0)}$, $c_j^{(k)}$ are given by
 \beq
 r^{(0)}(\xi,t) = r^{(0)}(\xi,0) e^{8 \i \xi^3 t},\hspace{5mm}
 r^{(k)}(\xi,t) = r^{(k)}(\xi,0) e^{8 \i \xi^3 t},
 \label{r_time}
 \eeq
 \beq
 c_j^{(0)}(t) = c_j^{(0)}(0) e^{8 \i \z_j^3 t},\hspace{5mm}
 c_j^{(k)}(t) = c_j^{(k)}(0) e^{8 \i \z_j^3 t}.
 \label{c_time}
 \eeq
It should be noted that $F(x, t)$ and $\5{F}(x,t)$ are expressed as
\beq
F(x,t)= f^{(0)}(x,t)\eins + \sum_{k=1}^{2m-1}f^{(k)}(x,t) e_k,
\label{}
\eeq
\beq
\5{F}(x,t)= f^{(0)}(x,t)\eins - \sum_{k=1}^{2m-1}f^{(k)}(x,t) e_k,
\label{}
\eeq
where the real functions $f^{(0)}(x,t)$ and $f^{(k)}(x,t)$
 satisfy
\beq
(\6_t + \6_{xxx}) f^{(0)}(2x,t)=0, \hspace{5mm}
(\6_t + \6_{xxx}) f^{(k)}(2x,t)=0.
\label{}
\eeq
Taking into account the conditions (\ref{z_z})--(\ref{C_C}), 
we can advance the analysis in parallel with the discussion in \S 
\ref{ISM} for the matrix mKdV equation. 
Thus the initial value problem of the cmKdV equations 
can be solved by the ISM, as has been shown in \S \ref{ISM}.

We replace $N$ in \S \ref{ISM} by $2N$ and find 
the $N$-soliton solution of the cmKdV equations
 (\ref{CMKdV5}) with $ M =2m$ components,
 \bea
 Q^{(m)} (x,t)
&=& -2 K_1(x,x;t)
 \nn \\
 &=& -2 \i \sum_{k=1}^{2N} P_k (x,t) C_k(t)^{\dagger} 
  \e^{-2 \i \z_k^\ast x}
 \nn \\
 &=& -2 \i \, (\, \underbrace{\, I \; I \; \cdots \; I \,}_{2N} \, )
 \; S^{-1} \;
 \left(
 \begin{array}{c}
  C_1 (t)^{\dagger} \e^{-2\i \z_1^\ast x}   \\
  C_2 (t)^{\dagger} \e^{-2\i \z_2^\ast x}   \\
   \vdots  \\
  C_{2N} (t)^{\dagger} \e^{-2\i \z_{2N}^\ast x}   \\
 \end{array}
 \right),
 \label{c_N_soliton}
 \eea
where the matrix $S$ is given by
 \beq
 S_{lk} \equiv \de_{lk} I -\sum_{j=1}^{2N}
 \frac{\e^{2\i (\z_j -\z_l^{\ast})x}}
 {(\z_j -\z_k^{\ast})(\z_j-\z_l^{\ast})} C_l(t)^\dagger C_j(t),
 \hspace{5mm} 1 \le l,k \le 2N.
 \label{}
 \eeq
\label{}

It is not evident whether eq.\ 
(\ref{c_N_soliton}) can be expressed as 
 \beq
 Q^{(m)}(x, t)= v_0(x,t) \eins + \sum_{k=1}^{2m-1} v_k(x,t) e_k, 
 \label{Q_1_ei}
 \eeq
without using $e_i e_j$, $e_i e_j e_k$, etc. But, noting the fact 
that 
summations and products of real quaternions are real quaternions,
 this can be proved for $m=2$ (4-component cmKdV equations) by using 
Neumann-Liouville expansion (see Appendix B). It is an open problem 
to prove eq.\ (\ref{Q_1_ei}) for general $M=2m$.

\section{Concluding Remarks}
\label{Concluding Remarks}

In this paper, we have constructed an extension of the ISM 
to solve the matrix mKdV equation and the matrix NLS equation. 
We get the coupled mKdV equations (\ref{CMKdV3}) as a 
reduction of the matrix mKdV equation. Through the extension, we
 have shown that the coupled 
mKdV equations have an infinite number of conservation laws and 
multi-soliton solutions and that its initial value problem is solvable. 

The existence of conserved quantities for the coupled mKdV equations 
has been proved by Svinolupov in a different
approach.~\cite{Svinolupov} Iwao and Hirota 
 obtained Pfaffian representation of $N$-soliton solution for the model 
by means of 
so-called Hirota's method.~\cite{Iwao} We stress that the initial value
 problem of the coupled mKdV equations has been solved in the present
paper for the first time. In addition, it directly 
proves the complete integrability of the 
model. Our scheme enables us to construct more general solution than 
the already known solutions. 

We can transform the
 coupled mKdV equations to a new integrable 
coupled version 
of the Hirota equation, ~\cite{Hirota2} by changes of dependent
variables $v_j$ and independent variables $x$, $t$. The new coupled 
Hirota equations describe interactions among different modes 
in optical fibers and seem to be physically significant. 
Wide applicability of our extension of the ISM will be reported 
in the succeeding papers. Further, the problem
 of integrable boundaries at 
$x=0$ for the coupled mKdV equations~\cite{Hirota3} will be
 studied elsewhere. 

After completing writing the paper, the authors were informed by 
Hisakado that the similar Lax formulation was used
 in the work of Eichenherr and Pohlmeyer.~\cite{EP}

\section*{Acknowledgement}
\setcounter{equation}{0}
One of the authors (T.~T.~) would like to express his sincere thanks to
Dr. M. Hisakado, Dr. M. Shiroishi, Dr. H. Ujino, Dr. N. Sasa and 
Mr.\ 
M.\ Ishikawa for many valuable comments and helpful discussions. 
In order to check the conservation laws
 of the coupled mKdV equations, the authors partly utilized useful 
{\it Mathematica} algorithm coded by \"{U}.\ G\"{o}kta\c{s} and W.\
 Hereman.~\cite{Goktas} He also 
appreciates the hospitality at Department of Applied Mathematics, 
University of Colorado, Boulder where some part of the work was done. 

\appendix
\section{Proof of eq.\ (\ref{det_A2})}
We begin with a counterpart of the linear equations
 (\ref{scattering_problem1}),
\beq
\Psi_x = U \Psi,
\label{}
\eeq
where $\Psi$ is assumed to be a square matrix. 
Using this equation, we get a chain of identities,
\bea
& \hspace{2mm} &
\Psi_x \Psi^{-1} = U,
\hspace{5mm}
\tr (\log \Psi)_x = \tr \hspace{0.5mm}U,
\nn \\
& \hspace{2mm} &
(\log \det \Psi)_x = \tr \hspace{0.5mm}U,
\nn \\
& \hspace{2mm} &
\det \Psi = \det \Psi(x_0) \cdot 
\e^{\tr \int_{x_0}^x \hspace{0.5mm}U \d x}.
\label{det_Psi}
\eea
In case that $U$ is given by 
\beq
 U
 =
 \left[
 \begin{array}{cc}
  -\i \z I  &  Q \\
   R  &  \i \z I \\
 \end{array}
 \right],
\label{}
\eeq
with square matrices $I$, $Q$ and $R$, eq.\ (\ref{det_Psi}) leads to
\beq
\det \Psi = {\rm const}.
\label{}
\eeq
If we take $[ \hspace{1mm}\5{\phi} \hspace{2mm} \5{\psi}
\hspace{1mm}]$ as $\Psi$, we get
\beq
\det [ \hspace{1mm}\5{\phi} \hspace{2mm} \5{\psi} \hspace{1mm}]
 = \det A^{\dagger}(\z^{\ast})
 = \det \5{A}(\z),
\label{}
\eeq
that means,
\beq
\det \5{A}(\z) = \{\det A (\z^{\ast})\}^{\ast}.
\label{}
\eeq

\section{Neumann-Liouville Expansion}
Due to the Gel'fand-Levitan-Marchenko equations (\ref{GL1}) and 
(\ref{GL2}), we get closed integral equations for $K_1$, $\5{K}_2$ 
and $K_2$, $\5{K}_1$,
\beq
K_1 (x,y) = \5{F}(x+y) - \int_x^{\infty} 
\d s_1 \int_x ^{\infty} \d s_2 K_1 (x, s_2) F(s_2 + s_1) \5{F}(s_1 +y),
\label{}
\eeq
\beq
\5{K}_2 (x,y) = -F(x+y) - \int_x^{\infty} 
\d s_1 \int_x ^{\infty} \d s_2 \5{K}_2 (x, s_2) \5{F}(s_2 + s_1) 
F(s_1 +y),
\label{}
\eeq
\beq
K_2 (x,y) = -\int_x^{\infty}\d s F(x+s)\5{F}(s+y) - \int_x^{\infty} 
\d s_1 \int_x ^{\infty} \d s_2 K_2 (x, s_2) F(s_2 + s_1) \5{F}(s_1 +y),
\label{}
\eeq
\beq
\5{K}_1 (x,y) = -\int_x^{\infty}\d s \5{F}(x+s)F(s+y) - \int_x^{\infty} 
\d s_1 \int_x ^{\infty} \d s_2 \5{K}_1 (x, s_2) \5{F}(s_2 + s_1) 
F(s_1 +y).
\label{}
\eeq
By successive approximations, we obtain the 
Neumann-Liouville expansions for $K_1$, $\5{K}_2$ 
and $K_2$, $\5{K}_1$,
\bea
K_1 (x,x) &=& \5{F}(2x) 
\nn \\
&& - \int_x ^{\infty} 
\d s_1 \int_x ^{\infty} \d s_2 \5{F} (x+s_2) F(s_2 + s_1) \5{F}(s_1 +x)
\nn \\
&& + \cdots 
\nn \\
&& + (-1)^n \int_x ^{\infty} \d s_1 \int_x^{\infty} \d s_2 
\cdots \int_x ^{\infty} \d s_{2n} \5{F}(x+s_{2n}) F(s_{2n} + s_{2n-1})
\5{F}(s_{2n-1} +s_{2n-2}) 
\nn \\
&& \hspace{50mm} \cdots F (s_2+s_1) \5{F}(s_1 +x)
\nn \\
&& + \cdots,
\label{}
\eea
\bea
\5{K}_2 (x,x) &=& -F(2x) 
\nn \\
&& + \int_x ^{\infty} 
\d s_1 \int_x ^{\infty} \d s_2 F(x+s_2) \5{F}(s_2 + s_1) F(s_1 +x)
\nn \\
&& + \cdots 
\nn \\
&& + (-1)^{n-1} \int_x ^{\infty} \d s_1 \int_x ^{\infty} \d s_2 
\cdots \int_x ^{\infty} \d s_{2n} F(x+s_{2n}) \5{F}(s_{2n} + s_{2n-1})
F(s_{2n-1} +s_{2n-2}) 
\nn \\
&& \hspace{50mm} \cdots \5{F} (s_2+s_1) F(s_1 +x)
\nn \\
&& +\cdots,
\label{}
\eea
\bea
K_2 (x,x) &=& -\int_x^{\infty} \d s_1 F(x+s_1) \5{F}(s_1 +x)
\nn \\
&& + \int_x ^{\infty} \d s_1 \int_x ^{\infty} \d s_2 
\int_x ^{\infty} \d s_3 F(x+s_3) \5{F}(s_3 + s_2) F(s_2 +s_1)
\5{F}(s_1 +x)
\nn \\
&& + \cdots 
\nn \\
&& + (-1)^{n-1} \int_x ^{\infty} \d s_1 \int_x ^{\infty} \d s_2 
\cdots \int_x ^{\infty} \d s_{2n+1} F(x+s_{2n+1}) \5{F}(s_{2n+1} + s_{2n})
F(s_{2n} +s_{2n-1}) 
\nn \\
&& \hspace{50mm} \cdots F(s_2+s_1) \5{F}(s_1 +x)
\nn \\
&& + \cdots,
\label{}
\eea
\bea
\5{K}_1 (x,x) &=& -\int_x^{\infty} \d s_1 \5{F}(x+s_1) F(s_1 +x)
\nn \\
&& + \int_x ^{\infty} \d s_1 \int_x ^{\infty} \d s_2 
\int_x ^{\infty} \d s_3 \5{F}(x+s_3) F(s_3 + s_2) \5{F}(s_2 +s_1)
F(s_1 +x)
\nn \\
&& + \cdots 
\nn \\
&& + (-1)^{n-1} \int_x ^{\infty} \d s_1 \int_x ^{\infty} \d s_2 
\cdots \int_x ^{\infty} \d s_{2n+1} \5{F}(x+s_{2n+1}) F(s_{2n+1} + s_{2n})
\5{F}(s_{2n} +s_{2n-1}) 
\nn \\
&& \hspace{50mm} \cdots \5{F}(s_2+s_1) F(s_1 +x)
\nn \\
&& + \cdots .
\label{}
\eea

\section{Proof of Proposition 4.1}
%\section{Proof of eqs.\ (\ref{r_r}) and (\ref{C_C})}

We remember $\Gamma = \Psi_2 \Psi_1^{-1}$ defined 
in \S \ref{Conservation laws}. The relation (\ref{cons4}) with 
eq.\ (\ref{cons5}) becomes
\beq
\Gamma_x = 2\i \z \Gamma + R^{(m)} - \Gamma Q^{(m)} \Gamma ,
\label{gam1}
\eeq
for the cmKdV equations. We assume 
\beq
\lim_{|\z| \to \infty}\Gamma = O \hspace{2mm} ({\rm Im}\, \z > 0),
\label{}
\eeq
and expand $\Gamma$ as
\beq
\Gamma = \sum_{l=1}^{\infty} \frac{1}{(2\i \z)^l} G_l,
\label{gam2}
\eeq
instead of eq.\ (\ref{cons8}). 
Substituting eq.\ (\ref{gam2}) for eq.\ (\ref{gam1}), we obtain 
a recursion formula,
\beq
G_{l+1} = -\de_{l,0} R^{(m)} + (G_l)_x + \sum_{j=1}^{l-1}
G_j Q^{(m)} G_{l-j}, \hspace{3mm} l=0,1,\cdots,
\label{gam3}
\eeq
where $Q^{(m)}$ and $R^{(m)}$ are given by
\beq
 Q^{(m)}= v_0 \eins + \sum_{k=1}^{2m-1} v_k e_k, 
 \hspace{5mm}
 R^{(m)}= -v_0 \eins + \sum_{k=1}^{2m-1} v_k e_k.
\label{}
\eeq
We first show the following theorem. \\
{\bf Theorem C. 1}
 $\hspace{5mm}$ 
Let $X$, $Y$, $Z$ be $2^{m-1} \times 2^{m-1}$ matrices 
given by
\bseq
\beq
 X= x_0 \eins + \sum_{k=1}^{2m-1} x_k e_k, 
\label{}
\eeq
\beq
 Y= y_0 \eins + \sum_{k=1}^{2m-1} y_k e_k, 
\label{}
\eeq
\beq
 Z= z_0 \eins + \sum_{k=1}^{2m-1} z_k e_k, 
\label{}
\eeq
\eseq
where the coefficients 
$x_0$, $x_k$; $y_0$, $y_k$; $z_0$, $z_k$ are real. 
Then there exist real numbers $w_0$, $w_k$ that satisfy 
\beq
W \equiv XYZ + ZYX = w_0 \eins + \sum_{k=1}^{2m-1} w_k e_k.
\label{W_eq}
\eeq
{\bf Proof} $\hspace{5mm}$  By the use of eq.\ (\ref{ei_ej}),
 a direct calculation gives
\bea
W &=& \biggl(x_0 \eins + \sum_{i=1}^{2m-1} x_i e_i \biggr)
 \biggl(y_0 \eins + \sum_{j=1}^{2m-1} y_j e_j \biggr)
 \biggl( z_0 \eins + \sum_{k=1}^{2m-1} z_k e_k \biggr)
\nn \\
&& +\biggl(z_0 \eins + \sum_{k=1}^{2m-1} z_k e_k \biggr)
  \biggl(y_0 \eins + \sum_{j=1}^{2m-1} y_j e_j\biggr)
  \biggl(x_0 \eins + \sum_{i=1}^{2m-1} x_i e_i\biggr)
\nn \\
&=&
 2 x_0 y_0 z_0 \eins 
 +2 x_0 y_0 \sum_{k=1}^{2m-1} z_k e_k 
 +2 x_0 z_0 \sum_{j=1}^{2m-1} y_j e_j
 +2 y_0 z_0 \sum_{i=1}^{2m-1} x_i e_i  
\nn \\
&& 
 +x_0 \biggl\{ \sum_{j,k=1}^{2m-1} y_j z_k (e_j e_k + e_k e_j)\biggr\}
 +y_0 \biggl\{ \sum_{i,k=1}^{2m-1} x_i z_k (e_i e_k + e_k e_i)\biggr\}
 +z_0 \biggl\{ \sum_{i,j=1}^{2m-1} x_i y_j (e_i e_j + e_j e_i)\biggr\}
\nn \\
&&
 + \sum_{i,j,k=1}^{2m-1} x_i y_j z_k (e_i e_j e_k + e_k e_j e_i)
\nn \\
&=& 
 2 x_0 y_0 z_0 \eins 
 +2 x_0 y_0 \sum_{k=1}^{2m-1} z_k e_k 
 +2 x_0 z_0 \sum_{j=1}^{2m-1} y_j e_j
 +2 y_0 z_0 \sum_{i=1}^{2m-1} x_i e_i  
\nn \\
&& 
 -2 x_0 \sum_{i=1}^{2m-1}y_i z_i \eins
 -2 y_0 \sum_{i=1}^{2m-1}x_i z_i \eins
 -2 z_0 \sum_{i=1}^{2m-1}x_i y_i \eins
\nn \\
&&
 -2 \sum_{i=1}^{2m-1}x_i y_i \sum_{k=1}^{2m-1} z_k e_k
 +2 \sum_{k=1}^{2m-1}x_k z_k \sum_{j=1}^{2m-1} y_j e_j
 -2 \sum_{j=1}^{2m-1}y_j z_j \sum_{i=1}^{2m-1} x_i e_i.
\label{}
\eea
This result shows that $W$ does not include terms like 
$e_i e_j$ or $e_i e_j e_k$ and can be expressed in the form of eq.\ 
(\ref{W_eq}). $\Box$
\\
Using Theorem C. 1, we can prove by the inductive method that 
$\{ G_l \}$ are expressed by
\beq
G_l = g_l^{(0)} \eins + \sum_{k=1}^{2m-1} g_l^{(k)} e_k,
\label{gam4}
\eeq
where $g_l^{(0)}$ and $g_l^{(k)}$ are real coefficients. 
Therefore, $\Gamma$ is given by
\bea
\Gamma &=& \sum_{l=1}^{\infty}\frac{1}{(2\i \z)^l} 
   \biggl( g_l^{(0)} \eins + \sum_{k=1}^{2m-1} g_l^{(k)} e_k\biggr)
\nn \\
&=& \sum_{l=1}^{\infty} \frac{g_l^{(0)}}{(2\i \z)^l} \eins
   + \sum_{k=1}^{2m-1} \biggl( 
        \sum_{l=1}^{\infty} \frac{g_l^{(k)}}{(2\i \z)^l}
        \biggr) e_k
\nn \\
&=& \gamma^{(0)}(\z) \eins + \sum_{k=1}^{2m-1}\gamma^{(k)}(\z)e_k,
\label{gam5}
\eea
where $\gamma^{(0)}(\z)$ and $\gamma^{(k)}(\z)$ satisfy
\beq
\gamma^{(0)}(\z) = \{ \gamma^{(0)}(-\z^{\ast})\}^{\ast},
\hspace{5mm}
\gamma^{(k)}(\z) = \{ \gamma^{(k)}(-\z^{\ast})\}^{\ast}.
\label{gam6}
\eeq
We recall the asymptotic behavior of the Jost function $\phi$ at
 $x \to \pm \infty$,
 \bea
 \phi &\sim&
%\left\{
%\begin{array}{c}
 \left[
 \begin{array}{c}
   I  \\
   O  \\
 \end{array}
 \right]
 \e^{-\i \z x}
 \hspace{7mm}
 {\rm as}~~~ x \rightarrow -\infty,
\\
\hspace{15mm}
&\sim& \left[
 \begin{array}{c}
   A(\z)  \e^{-\i \z x}  \\
   B(\z)  \e^{\i \z x} \\
 \end{array}
 \right]
 \hspace{7mm}
 {\rm as}~~~ x \rightarrow +\infty.
%\end{array}
%\right.
 \label{}
\eea
These relations yield
\beq
\lim_{x \to - \infty} \phi_1 \e^{\i \z x} = I, \hspace{2mm}
\lim_{x \to + \infty} \phi_1 \e^{\i \z x} = A(\z),
\label{}
\eeq
where we have defined $\phi_1$ and $\phi_2$ by
\beq
 \phi =
 \left[
 \begin{array}{c}
   \phi_1  \\
   \phi_2  \\
 \end{array}
 \right].
\label{}
\eeq
We easily see that
\beq
\lim_{|\z| \to \infty}\phi_2 \phi_1^{-1}
 = O \hspace{2mm} ({\rm Im}\, \z > 0),
\label{}
\eeq
then we can replace $\Gamma$ in eq.\ (\ref{gam5}) with 
$\phi_2 \phi_1^{-1}$. Thus $\det A(\z)$ is expressed as
\bea
 \det A(\z) 
&=& \exp\{\tr \log A(\z)\}
\nn \\
&=& \exp \biggl\{  \tr \int_{-\infty}^\infty (\log 
        \phi_1 \e^{\i \z x})_x \d x \biggr\}
\nn \\
  &=& \exp \biggl\{  \tr \int_{-\infty}^\infty 
        (\phi_1 \e^{\i \z x})_x (\phi_1 \e^{\i \z x})^{-1}
        \d x \biggr\}
\nn \\
  &=& \exp \biggl\{  \tr \int_{-\infty}^\infty 
        Q^{(m)} \phi_2 \phi_1^{-1} \d x \biggr\}
\nn \\
  &=& \exp \biggl\{  \tr \int_{-\infty}^\infty 
        \biggl( v_0 \gamma^{(0)} (\z) - \sum_{k=1}^{2m-1} 
        v_k \gamma^{(k)}(\z)\biggl) \eins \, \d x \biggr\}
\nn \\
  &=& \exp \biggl\{  2^{m-1}\int_{-\infty}^\infty 
        \biggl( v_0 \gamma^{(0)} (\z) - \sum_{k=1}^{2m-1} 
        v_k \gamma^{(k)} (\z)\biggr) \d x \biggr\},
\label{}
\eea
where we have used eqs.\ (\ref{tr_ei}) and (\ref{tr_ei_ej}). 
Due to eq.\ (\ref{gam6}), $\det A(\z)$, as a function of $\z$,
 satisfies
\beq
\det A(\z) = \{ \det A(-\z^{\ast}) \}^{\ast}.
\label{}
\eeq
This is the proof of Proposition 4.1$\;$(1). Further, we obtain
\bea
B(\z) A(\z)^{-1} &=& \lim_{x \to +\infty} \phi_2 \phi_1^{-1}
        \e^{-2\i \z x}
\nn \\
&=& \lim_{x \to +\infty} \biggl[ \gamma^{(0)} (\z) \e^{-2\i \z x} \eins
        + \sum_{k=1}^{2m-1} \gamma^{(k)} (\z) \e^{-2\i \z x} e_k \biggr]
\nn \\
&=&  r^{(0)}(\z) \eins + \sum_{k=1}^{2m-1} r^{(k)}(\z) e_k,
\label{BAr1}
\eea
with conditions
\beq
r^{(0)}(\z) = \{r^{(0)}(-\z^{\ast})\}^{\ast},
\hspace{2mm} r^{(k)}(\z) = \{r^{(k)}(-\z^{\ast})\}^{\ast}.
\label{BAr2}
\eeq
Using eqs.\ (\ref{BAr1}) and (\ref{BAr2}), 
it is straightforward to prove Proposition 4.1$\;$(2)(3).

%\section{Proof of eq.\ (\ref{})}

%
{\it Note added in proof}$\,$---$\,$Equation (\ref{c_N_soliton}) 
includes not only pure soliton solutions but also breather solutions. 
We should impose appropriate conditions on the arbitrary parameters in 
the residue matrices, {\it e.g.},
$C_{2j-1}\bar{C}_{2j}=\bar{C}_{2j}C_{2j-1}
=C_{2j}\bar{C}_{2j-1}=\bar{C}_{2j-1}C_{2j}=O$, or equivalently
\[
\sum_{i=0}^{2m-1}(c_j^{(i)})^2 =0,
\]
to obtain pure soliton solutions. In this case, the 1-soliton solution 
for the cmKdV equations (\ref{CMKdV5}) is given by
\bea
Q^{(m)}(x,t) 
&=& 2\eta_1 \sech \{2\eta_1 x -8\eta_1 (\eta_1^2-3\xi_1^2)t-x_0 \} 
    \biggl( 2 \sum_{l=0}^{2m-1} |c_1^{(l)}|^2 \biggr)^{-\frac{1}{2}}
\nn \\
&& \cdot \{ -\i \bar{C}_1 (0) \e^{-2\i \xi_1 x -8\i \xi_1 
 (\xi_1^2-3\eta_1^2)t} -\i \bar{C}_2 (0) \e^{2\i \xi_1 x + 8\i \xi_1 
 (\xi_1^2-3\eta_1^2)t} 
 \},
\nn
\eea
where $x_0$ is defined by 
\[
\e^{-x_0} = 2\eta_1 \biggl( 2 \sum_{l=0}^{2m-1} 
  |c_1^{(l)}|^2  \biggr)^{-\frac{1}{2}} \, .
\]
For instance, $v_0 (x,t)$ (cf. (\ref{CMKdV5})) is given by
\bea
v_0(x,t)
&=& 2 \eta_1 \sech \{ 2 \eta_1 x - 8 \eta_1 (\eta_1^2-3\xi_1^2)t-x_0\}
\biggl( 2 \sum_{l=0}^{2m-1} |c_1^{(l)}|^2 \biggr)^{-\frac{1}{2}}
\nn \\
&& 
\cdot \{c_1^{(0) \ast } \e^{-2\i \xi_1 x-8\i \xi_1 (\xi_1^2-3\eta_1^2)t}
 + c_1^{(0)} \e^{2\i \xi_1 x+8\i \xi_1 (\xi_1^2-3\eta_1^2)t} \}.
\nn 
\eea

\end{document}